\documentclass[12pt]{article}

\usepackage{graphicx}
\usepackage{amsmath}
\usepackage{amssymb}
\usepackage{amsbsy}
\usepackage{mathtools}
\usepackage[left, running, displaymath, mathlines]{lineno}
 \usepackage{mathptmx}           % this gives slanted Times symbols for math
\usepackage{bm}   % for bold math fonts, especially greek
\usepackage{framed}  % for frame boxes
\numberwithin{equation}{section}
\newcommand{\bit}{\begin{itemize}}
\newcommand{\eit}{\end{itemize}}

\def\benu{\begin{enumerate}}
\def\eenu{\end{enumerate}}
\def\noi{\noindent}
\def\btab{\begin{tabbing}}
\def\etab{\end{tabbing}}

\def\bit{\begin{itemize}}
\def\eit{\end{itemize}}
\def\beq{\begin{equation}}
\def\eeq{\end{equation}}
\def\bec{\begin{center}}
\def\eec{\end{center}}
\def\btable{\begin{tabular}}
\def\etable{\end{tabular}}
\def\beqr{\begin{eqnarray}}
\def\eeqr{\end{eqnarray}}
\def\rarw{\rightarrow}
\def\Rarw{\Rightarrow}

\def\gm{\gamma}

\def\lm{\lambda}
\def\eps{\epsilon}

\def\al{\alpha}
\def\bt{\beta}

\def\dl{\delta}
\def\Dl{\Delta}
\def\sg{\sigma}

\def\rarw{\rightarrow}
\def\del{\partial}

\def\half{\frac{1}{2}}

\def\ns{\normalsize}
\def\btab{\begin{tabbing}}
\def\etab{\end{tabbing}}
\def\beqrs{\begin{eqnarray*}}
\def\eeqrs{\end{eqnarray*}}
\def\noi{\noindent}
\def\lan{\langle}
\def\ran{\rangle}
\def\bibi{\bibitem}
\def\bfig{\begin{figure}}
\def\efig{\end{figure}}
\def\fr{\frac}
\def\non{\nonumber}
\def\barr{\begin{array}}
\def\earr{\end{array}}

 \fontfamily{ptm}\selectfont
\title{\mbox{}  \hfill  {\normalsize FERMILAB-PUB-23-279-AD} \\
  \mbox{} \\ 
{\bf \ns Density distributions of tune shifts from space charge or beam-beam interactions in Gaussian bunches}
}
{\ns 
\author{Tanaji Sen  \footnote{tsen@fnal.gov} \\
Fermi National Accelerator Laboratory, Batavia, IL 60510 }
}
\date{}

\begin{document}

\maketitle

\begin{abstract}
  The amplitude dependent tune shifts  from either space
  charge or beam-beam interactions are calculated analytically  with the inclusion of synchrotron
  oscillations and multiple interactions around the ring. Simpler formulae are derived under limits  of bunches
  longer than the transverse sizes,
  equal and unequal transverse sizes etc. This is used to derive semi-analytical forms for the density distribution
  of the tune shifts. The tune spread and  the density distribution are needed to understand beam decoherence
  or Landau damping with either interaction.   The tune footprints due to space charge in IOTA are simulated
  using pyorbit and found to be in good agreement with the theoretical predictions. 

\end{abstract}

\section{Introduction}

The space charge interaction in a low energy synchrotron and  the beam-beam
interaction in a collider are the dominant contributors to the incoherent tune spread in these machines.
In this report we calculate first the incoherent  amplitude dependent  transverse tune shifts in
Gaussian beams due to either interaction.
We generalize these tube shifts to include the effects of synchrotron oscillations and especially in the case of
space charge, we also include the contributions of the interactions from multiple locations around the ring.
Next we calculate the beam density distributions as a function of these tuneshifts. This density
distribution is needed to determine beam stability in different conditions. In terms of scaled tuneshifts
(defined in Section \ref{Sec: density}), the density distribution has exactly the same form for both space charge and
beam-beam interactions. However, the role of the tune spread and the density in determining beam stability
is very different in the two interactions. The beam-beam interactions act as an external source of tune spread
and can consequently be used to provide Landau damping \cite{Buffat}
while with space charge, an external driving source such as octupoles \cite{Landau_oct} or perhaps
electron lenses \cite{Landau_elens} is required for Landau damping. Nevertheless even with space charge, the 
contributions from the internal tune spread and the density distribution have to be included in determining
beam stability.  Another use of the incoherent tune spread with either interaction is finding
the beam decoherence time when the centroid is offset from the center e.g. due to a  dipole kick.

. The  calculation of tune shifts due to head-on beam-beam interactions in one dimension 
was reported in \cite{Keil_94} while the fully 2D calculation of the tune spreads, resonance driving terms
etc was done e.g in \cite{Sen_2003}. These were then  generalized to long-range interactions
\cite{Sen_PRAB03} which were of greater interest in the Tevatron. The expressions for head-on interactions are
easily found by taking the limit of zero separation.
We note that the space charge tune shifts with amplitude for round beams
without synchrotron oscillations were calculated in \cite{Ng_2004} which used some of the methods in
\cite{Sen_2003}. The density distribution was extracted from numerical simulations, but insufficient
sampling of the beam core led to an incorrect form for the density, especially close to the core where the
space charge tune shift is largest. Our method is semi-analytical, in that numerical inversion of analytical
functions followed by interpolation to obtain smooth functions is required. We also check that the zeroth to
second moments of the distribution are preserved.

\section{Incoherent tune shifts with synchrotron oscillations} \label{Sec: tunes_theory}

Here we consider the tune shifts with amplitude due to a Hamiltonian with linear transverse motion,
longitudinal motion in an rf bucket and either a space charge interaction or
a beam-beam interaction. In  this section, we consider first the Hamiltonian with
a space charge interaction experienced by a Gaussian distribution in three space dimensions.
At the end of this section, we consider the beam-beam interaction and show that the tuneshift with
amplitude scaled by the zero amplitude tune shift has the same form as with space charge.
The  Hamiltonian in the lab frame can be written in dimensionless form as
\beq
H = \half( (x')^2 + (y')^2 + K_x x^2 + K_y y^2 ) + \fr{e}{\bt^2 \gm m_0 c^2} ( V_{rf} + V_{sc})
\eeq
where $V_{rf}$ is the  rf cavity voltage wave form and $V_{SC}$ is the electric potential  due to  the
space charge measured in the lab frame.   Transforming to action-angle variables $(J_x, \phi_x, J_y, \phi_y)$
in the transverse planes, as e.g.
\beqr
x & = & \sqrt{2 \bt_x J_x}\cos \phi_x,  \;\;\; x' =  \sqrt{2 J_x}[\sqrt{\fr{1}{\bt_x}} \sin \phi_x - 2\al_x\cos\phi_x]
\eeqr
This reduces the linear part of the transverse Hamiltonian to
\beq
H_{\perp, 0} = \fr{1}{R}(\nu_{x, 0} J_x + \nu_{y, 0}  J_y)
\eeq
where $(\nu_{x, 0}, \nu_{y, 0})$ are the tunes of the linear lattice and $R$ is the machine radius.

Consider a Gaussian distribution in 3 space dimensions for a bunch; the same bunch experiencing its space charge
field or
the opposing bunch for the case of beam-beam interactions
\beq
\psi(x, y, z)  = \fr{N e}{(2 \pi)^{3/2} \sg_x \sg_y \sg_z}\exp[ - \fr{x^2}{2\sg_x^2} - \fr{y^2}{2\sg_y^2}
- \fr{z^2}{2\sg_z^2} ]
\eeq
where $\sg_x, \sg_y, \sg_z$ are the rms bunch dimensions. The solution of Poisson's equation
$\nabla^2V = - \psi/\eps_0$ leads to the following solution for the electric scalar potential \cite{Takayama_80}
\beqr
V(x, y, z)  & =  & \fr{1}{4\pi \eps_0}\fr{N e}{\pi^{1/2} \gm}\int_0^{\infty} dq \;
\fr{1}{\sqrt{(2\sg_x^2 + q)(2\sg_y^2 + q)(2\gm^2 \sg_z^2 + q) }} \non \\
&     &
\left[1 - \exp\left[(- \fr{x^2}{2\sg_x^2 + q} - \fr{y^2}{2\sg_y^2 + q}- \fr{\gm^2 z^2}{2\gm^2 \sg_z^2 + q}
    \right)   \right] 
\eeqr
where the coordinates $(x, y, z)$ and the rms sizes $(\sg_x, \sg_y, \sg_z)$ are measured in the rest frame. 
The complete Hamiltonian in three degrees of freedom (3D) after scaling by $R$ is
\beqr
H  &  =  & \nu_{x, 0} J_x + \nu_{y, 0}J_y + \fr{e R}{\bt^2 \gm m_0 c^2} V_{rf}(\dl p/p, z)  +
C_{SC} \bar{V} (x, y, z) \\
C_{SC} & =   & \fr{N_p r_p}{\pi^{1/2}\bt^2 \gm^2}  \\
  \bar{V} (x, y, z) & = & \int_0^{\infty} dq \; \fr{1}{\sqrt{(2\sg_x^2 + q)(2\sg_y^2 + q)(2\gm^2 \sg_z^2 + q) }} \non \\
&     & \left[1 - \exp(- \fr{x^2}{2\sg_x^2 + q} - \fr{y^2}{2\sg_y^2 + q}- \fr{\gm^2 z^2}{2\gm^2 \sg_z^2 + q} )  \right]
\eeqr
where $r_p = e^2/(4\pi \eps_0 m_0 c^2)$ is the classical particle radius. 
  The tunes follow from the derivatives of the angle-averaged Hamiltonian. 
We will write only the expressions in $x$, the one for $y$ can be found by the replacement $x \leftrightarrow y$.
Our focus is on the transverse tune shifts with amplitude, thus we ignore dominantly longitudinal effects such as
longitudinal space charge effects.  We also do not consider the momentum
dependence of the transverse tunes or the modulation of the revolution period by the synchrotron oscillations, these
do not have a noticeable effect on the dynamics in IOTA because of the small synchrotron tune.
  We do include the impact of synchrotron
  oscillations on the transverse dynamics via the nonlinear interaction potential. With these assumptions,  the transverse tune shifts are given by
  \beq
  \Dl \nu_x = R  C_{SC} \fr{\del}{\del J_x} \lan \bar{V} \ran_{\phi_x, \phi_y , \phi_z , s}
  \eeq
  where the averaging is done over all three angles and over $s$, the  length along the ring.
Hence after using action-angle variables, 
  \beqrs
  \Dl\nu_x & = & R C_{SC} \int_0^{\infty} dq \;
\fr{1}{\sqrt{(2\sg_x^2 + q)^3 (2\sg_y^2 + q)(2\gm^2 \sg_z^2 + q) }}
2 \bt_x \cos^2\phi_x \\
& & \exp \left[- \fr{ 2 \bt_x J_x \cos^2\phi_x}{2\sg_x^2 + q} - \fr{y^2}{2\sg_y^2 + q}- \fr{\gm^2 z^2}{2\gm^2 \sg_z^2 + q} \right]
   \eeqrs
Change the integration variable from $q$ to a dimensionless variable $u$ as 
\beqrs
u & = & \fr{2\sg_x^2}{(2\sg_x^2 + q)}, \;\;\; \Rarw q = \fr{2\sg_x^2}{u} - 2\sg_x^2
\eeqrs
This converts the infinite range of integration over $q$ to a finite range of integration over $u$. 
Hence
\beqr
\Dl\nu_x & = & \fr{ R C_{SC}  \bt_x }{2^{1/2} \sg_x^2 \gm \sg_z}  \int_0^1   du \; 
\Big\lan \cos^2\phi_x     \left[\fr{1}{[( \sg_y^2/\sg_x^2 - 1 )u + 1 ]}
    \fr{u}{[(1 - \sg_x^2/\gm^2\sg_z^2 )u + \sg_x^2/\gm^2\sg_z^2]} \right]^{1/2}  \non \\
&  &  \exp \left[- \fr{ 2 \bt_x J_x \cos^2\phi_x}{2\sg_x^2}u  - \fr{ 2 \bt_y J_y \cos^2\phi_y}{2\sg_y^2}
   \fr{u}{2\sg_y^2[(1 - \sg_x^2/\sg_y^2 )u + \sg_x^2/\sg_y^2]} \right.  \non \\
&  & \left.  -\fr{\gm^2 z^2 u}{2\gm^2\sg_z^2[(1 - \sg_x^2/\gm^2\sg_z^2 )u + \sg_x^2/\gm^2\sg_z^2]} \right] \ran_{\phi_x, \phi_y ,  s} 
\eeqr

In order to make progress, we need to assume that the longitudinal motion  is simple
harmonic. This implies that the rf cavity force be linear of equivalently that we approximate the cosine term in
$V_{rf}$ by the first two terms in  its Taylor expansion.  Writing the coordinates $(x, y, z)$  in terms of dimensionless amplitudes $(a_x, a_y, a_z)$ and the
corresponding  rms sizes $(\sg_x, \sg_y, \sg_z)$
  \beq
  x = a_x \sg_x \cos\phi_x , \;\;\;     y = a_y \sg_y \cos\phi_y , \;\;\;   z = a_z \sg_z \cos\phi_z
  \eeq
We  use the integral representation of $I_0$
\beq
I_0(w) = \fr{1}{\pi}\int_0^{\pi} d\theta \exp[\pm w \cos\theta] = \fr{1}{2\pi}\int_0^{2\pi} d\theta
\exp[\pm w \cos 2\theta]
\eeq
The integrals in the phase averages over $\phi_x, \phi_y, \phi_z$  are of the form
\beqrs
\fr{1}{2\pi}\int_0^{2\pi} d\phi  \exp[- w\cos^2\phi]   & = &    \exp[ -\half w] I_0(\half w) \\
\fr{1}{2\pi}\int_0^{2\pi} d\phi \cos^2\phi \exp[- w\cos^2\phi] & = &
\half \exp[ -\half w] [I_0(\half w) - I_1(\half w) ]  
\eeqrs
Gathering all terms together,  the final expression is
\beqr
\Dl\nu_x(a_x, a_y, a_z) & = & C_{SC} \frac{  R}{2 \sqrt{2} \gm \sg_z \eps_{x}} \int_0^1   du \;  
 \exp[- \fr{ a_x^2 u}{4}]\left[ I_0(\fr{ a_x^2 u}{4}) - I_1(\fr{ a_x^2 u}{4})\right]   
 \exp[- \fr{ a_y^2 u}{4}] \exp[- \fr{ a_z^2 u}{4}]   \non  \\
 &   & \times \Big\lan  \left[\fr{1}{[( \sg_y^2/\sg_x^2 - 1 )u + 1 ]}
   \fr{u}{[(1 - \sg_x^2/\gm^2\sg_z^2 )u + \sg_x^2/\gm^2\sg_z^2]} \right]^{1/2} \non  \\
 & & \times   I_0\left(\fr{ a_y^2}{4}\fr{u}{(1 - \sg_x^2/\sg_y^2 )u + \sg_x^2/\sg_y^2}\right)
 I_0\left(\fr{ a_z^2 u}{4[ (1 - \sg_x^2/\gm^2\sg_z^2 )u + \sg_x^2/\gm^2\sg_z^2 ] }\right)
 \Big\ran_s    \label{eq: dnux_gen}   \non \\
 \eeqr
From this general expression, we can obtain the tuneshifts for special cases.

\subsection{ Bunch length longer than the transverse sizes} \label{subsec: long}

Using the general expression Eq.(\ref{eq: dnux_gen}) in this limit where the transverse sizes are both
negligibly small compared to the bunch length, i.e. $(\sg_x, \sg_y) \ll \sg_z$
we find
\beqr
\Dl\nu_x(a_x, a_y, a_z) & = &    C_{SC} \frac{  R}{2 \sqrt{2} \gm \sg_z \eps_{x}} {\Big \lan}  
\int_0^1   du \;   \exp[- \fr{ a_z^2 u}{4}]  I_0\left(\fr{ a_z^2 u }{4 }\right)
 \exp[- \fr{ a_x^2 u}{4}]\left[ I_0(\fr{ a_x^2 u}{4}) - I_1(\fr{ a_x^2 u}{4})\right] 
 \non   \\
& & \times   \exp[- \fr{ a_y^2 u}{4}]  \left[\fr{1}{[( \sg_y^2/\sg_x^2 - 1 )u + 1 ]} \right]^{1/2}   
 I_0\left(\fr{ a_y^2}{4}\fr{u}{(1 - \sg_x^2/\sg_y^2 )u + \sg_x^2/\sg_y^2}\right)
{\Big  \ran_s}   \non \\
 &  &  \label{eq: dnux_long}
\eeqr
Eq.(\ref{eq: dnux_long}) and a similar one for $\Dl\nu_y$ (with $x \leftrightarrow y$) are of the form
which is generally applicable for bunches in hadron synchrotrons, even the low energy machines where
$\gm \simeq 1$.

The tuneshift at the origin  is
\beqr
\Dl \nu_{x, SC}(0, 0, 0)  & = &    \fr{ R C_{SC} }{2 \sqrt{2}\gm \eps_x \sg_z } \int_0^1   du \;
\non   \Big\lan  \left[\fr{1}{[( \sg_y^2/\sg_x^2 - 1 )u + 1 ]} \right]^{1/2}  \Big\ran_s  \non \\
& =  & \fr{ R C_{SC} }{ \sqrt{2}\gm  \eps_x \sg_z}\lan  \fr{1}{1 + \sg_y/\sg_x} \Big\ran_s
\eeqr
Substituting the expression for $C_{SC}$ and assuming round beams at all locations, we obtain 
\beqr
\Dl \nu_{x, SC} & =  & \fr{N_p r_p}{\bt^2 \gm^2} \fr{ R}{2\sqrt{2 \pi}\gm  \sg_z \eps_x } 
=\fr{r_p}{\bt\gm^2 \eps_{x, N}} \lm_G R , \;\;\;\; 
\lm_G  =  \fr{N_p }{2\sqrt{2 \pi} \sg_z}
\eeqr
where we used $\eps_{x, N} = \bt \gm \eps_x$ and $\lm_G$ is the longitudinal density. 
These are  the standard expressions for the space charge tune shift parameters for Gaussian bunches.
With a coasting bunch, the longitudinal density is $\lm = N_p/(2\pi R)$.

The zero amplitude tune shifts for non-round beams can be written as
\beqr
\Dl\nu_{x}(0, 0, 0) & = &  2 \Big\lan \fr{1}{ \sg_y/\sg_x + 1} \Big\ran_s \Dl\nu_{x, SC} , \;\;\;\;
\Dl\nu_{y}(0, 0, 0) = 2 \Big\lan \fr{1}{\sg_X/\sg_y + 1} \Big\ran_s \Dl\nu_{y, SC}  \label{eq: dq_nr_SC}
\eeqr
These equations include the variation of beam sizes around  the ring.

At zero transverse amplitude in this limit of both long and round bunches
\bfig  
\centering
\includegraphics[scale=0.5]{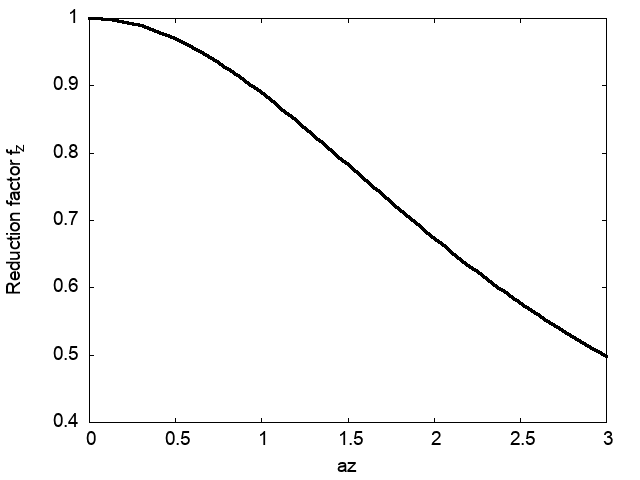}
\caption{Correction factor $f_z$ for the reduction of the zero amplitude tune shift as a function of the
 longitudinal amplitude $a_z$.}
\label{fig: facz}
\efig  
 \beqr
 \Dl\nu_x(0, 0, a_z)|_{Long, round} & = & \Dl\nu_{x, SC}    \int_0^1   du \;   \exp[- \fr{ a_z^2 u}{4}]  I_0\left(\fr{ a_z^2  u}{4 }\right)   \equiv    f_z \Dl\nu_{x, SC} \\
 f_z & =  & \exp[- \fr{ a_z^2 }{4}] \left( I_0(\fr{ a_z^2  }{4 }) +  I_1(\fr{ a_z^2  }{4 }) \right) 
 \eeqr
where $f_z$ is the correction factor which describes the impact of synchrotron oscillations.
Hence $f_z$ plotted in Fig. \ref{fig: facz}  shows that the small amplitude tune shifts due to space charge of a longitudinal slice decreases with distance $z$, falling by half at $a_z = 3$ corresponding to the
edge of the bucket. The caveat is that the longitudinal motion is not simple harmonic; nevertheless this curve
should be accurate at smaller amplitudes.
The small transverse amplitude particles can oscillate over different synchrotron
amplitudes; we can average over the full range of these amplitudes. Assuming a Gaussian distribution in $a_z$, we
find
\beq
\lan f_z \ran = \fr{\sqrt{2} (\pi^2 - 2 \Gamma[3/4]^4)}{\pi^{3/2} \Gamma[3/4]^2} \simeq 0.91
\eeq
We find that synchrotron oscillation reduces the small amplitude tune shift by about 10\%.
However since synchrotron oscillations have a much longer time scale than betatron oscillations, this average
value $\lan f_z \ran$ may not be a useful indicator of their impact. 

 Zero synchrotron amplitude  $a_z = 0$ with arbitrary transverse sizes
\beqr
\Dl\nu_x(a_x, a_y, 0)|_{Long} & = & \Dl\nu_{x, SC} {\Big \lan}    \int_0^1   du \;
 \exp[- \fr{ a_x^2 u}{4}]\left[ I_0(\fr{ a_x^2 u}{4}) - I_1(\fr{ a_x^2 u}{4})\right]   \non   \\
& & \times \exp[- \fr{ a_y^2 u}{4}] \left[\fr{1}{[( \sg_y^2/\sg_x^2 - 1 )u + 1 ]}
   I_0\left(\fr{ a_y^2}{4}\fr{u}{(1 - \sg_x^2/\sg_y^2 )u + \sg_x^2/\sg_y^2}\right) \right]
{\Big \ran_s   } \non \\ 
\mbox{} \label{eq: dq_Long_NR}
\eeqr
  We observe that when the beams are not round, the tuneshift depends on the variation of the relative
  beam size $\sg_y/\sg_x$ around the ring. The average over the ring can be calculated exactly with an
  integral evaluation at   several points along the ring or approximately by
  replacing the average over the function by the function of   the averaged argument.
 This approach would require a single integral and would be computationally faster. The quality of this
  approximation will be evaluated in  Section \ref{Sec: IOTA} for IOTA parameters.

With the further assumption of round bunches everywhere, the above simplifies to 
\beqr
\Dl\nu_x(a_x, a_y, a_z)|_{Long, round} & = &   \Dl\nu_{x, SC}
   \int_0^1   du \;   \exp[- \fr{ a_z^2 u}{4}]  I_0\left(\fr{ a_z^2  u}{4 }\right) \non   \\
& & \times  \exp[- \fr{ a_x^2 u}{4}]\left[ I_0(\fr{ a_x^2 u}{4}) - I_1(\fr{ a_x^2 u}{4})\right] 
 \exp[- \fr{ a_y^2 u}{4}] I_0\left(\fr{ a_y^2}{4} u\right)  \non \\
 \label{eq: dnux_long_round}
 \eeqr
In most machines the bunch is not round everywhere in the ring, nevertheless 
 Eq.(\ref{eq: dnux_long_round}) can be used as a first approximation for the tuneshift with amplitude.

In Section \ref{Sec: IOTA} we evaluate the tune shifts for IOTA parameters and consider the impact of
synchrotron oscillations and the round beams approximation on the transverse tune shifts.

\subsection{Beam-beam tune shifts}

The beam-beam tune footprint can be found in a similar fashion as above. The major difference is that in a
collider, the particles all move at very relativistic speeds, so $\bt \simeq 1, \gm \gg 1$. Consequently there is
a large increase in the transverse fields in going from the rest frame to the lab frame due to the Lorentz
boost while the longitudinal fields are unchanged.
\beq
    {\bf E}_{\perp, lab} \simeq \gm {\bf E}_{\perp, rest} = - \gm {\bf \nabla}_{\perp} V_{BB}, \;\;
    E_{z, lab} = E_{z, rest}, \;\;\;\; B_{x, lab}  = \fr{E_{y, lab}}{\bt}, \;\;\;\; B_{y, lab}  = -\fr{E_{x, lab}}{\bt}
    \eeq
    Here $V_{BB}$ is the scalar potential for the beam-beam interaction and is the same as $V_{SC}$ except that the
    beam parameters are of the opposing bunch. The net force due to the electric and magnetic fields are in the
    same direction for beam-beam interactions and oppose each other with space-charge. 
While the space charge forces act radially outward in all directions, the beam-beam forces are almost entirely
in the transverse plane emanating from a squashed pancake like disc traveling with the opposing beam. In
most circumstances we can think of the opposing beam as being point-like along the direction of motion and the beam-beam potential as effectively two dimensional.
The longitudinal density has a role to play in beam-beam interactions in effects such as phase
averaging in long bunches \cite{Krish_Sie, Sen_phase} or when hourglass effects \cite{Furman} or
crossing angles are introduced; see \cite{Sen_Higgs} for a recent
calculation  of the luminosity and beam-beam tune shifts with both these effects  in a Higgs factory
$e^+-e^-$ collider. 

In most cases where the beam-beam interaction is 2D, the results of section \ref{subsec: long} are
applicable here  because $\gm \gg 1 $.  Following the same procedure as in obtaining
Eq. (\ref{eq: dq_nr_SC})   leads to the beam-beam tune shift at the origin
\beq
\Dl \nu_{x, bb} = \fr{r_p N_p \bt_x^*}{2 \pi \gm}\fr{1}{\sg_x^*(\sg_x^* + \sg_y^*)}, \;\;\;
\Dl \nu_{y, bb} = \fr{r_p N_p \bt_y^*}{2 \pi \gm}\fr{1}{\sg_y^*(\sg_x^* + \sg_y^*)}
\eeq
where $\bt_x^*, \bt_y^*, \sg_x^*, \sg_y^*$ are the values at the IP and we assumed that the beam
parameters are the same  at all the IPs. 

The important point here  is that both the space charge and the beam-beam potential have the same
dependence on the transverse amplitudes, consequently the two footprints are the same if we scale out
the zero amplitude tune shifts.  Thus, the horizontal beam-beam tune shift at transverse amplitudes $(a_x, a_y)$  can be written down using Eq.(\ref{eq: dq_Long_NR}) 
\beqr
\Dl\nu_{x, bb}(a_x, a_y) & = & \Dl \nu_{x, bb}   \int_0^1   du \;
\exp[- \fr{ a_x^2 u}{4}]\left[ I_0(\fr{ a_x^2 u}{4}) - I_1(\fr{ a_x^2 u}{4})\right] \exp[- \fr{ a_y^2 u}{4}]
\non   \\
& & \times \left[\fr{1}{[( \sg_y^2/\sg_x^2 - 1 )u + 1 ]}
   I_0\left(\fr{ a_y^2}{4}\fr{u}{(1 - \sg_x^2/\sg_y^2 )u + \sg_x^2/\sg_y^2}\right) \right]
\eeqr
A similar expression holds in the vertical plane.

\section{Density distribution in tunes} \label{Sec: density}

We saw in the previous section that the beam and machine parameters describing the space charge and
beam-beam  tune
  shifts are all included in the zero amplitude tune shifts $ \Dl\nu_{x, sc},  \Dl\nu_{y, sc}$.  It is therefore
  useful to describe the universal functions of the dimensionless amplitudes as
  \beq
  \xi_x(a_x,  a_y, a_z) = \fr{\Dl\nu_x(a_x, a_y, a_z)}{\Dl\nu_{x, sc}}, \;\;\;\;
  \xi_y(a_x,  a_y, a_z) = \fr{\Dl\nu_y(a_x, a_y, a_z)}{\Dl\nu_{y, sc}}
  \eeq
The functions $\xi_x, \xi_y$ are universal in   the sense that their behavior describes the amplitude
  dependence for any machine. In this section we consider the distribution assuming a Gaussian distribution in phase space.

  \subsection{Density distribution in 1D}

We start with the tune density distribution in 1D for the sake of clarity. The density in action $j_x$ is
  transformed to the dimensionless amplitude variable $\al_x$ as
  \beqr
  \rho(j_x) & = & \fr{1}{\eps_x}\exp[- \fr{j_x}{\eps_x}] = \fr{1}{\eps_x}\exp[-2 \al_x] , \;\;\;
  \al_x = \fr{a_x^2}{4} = \fr{j_x}{2\eps_x} \\
 \rho(\al_x) & = & \rho(a_x) [\fr{\del \al_x}{\del a_x}]^{-1} =  a_x \exp[ - \half a_x^2 ] [ a_x/2 ]^{-1}
 = 2 \exp[- 2\al_x ]  
 \eeqr
Now  we need to transform from the amplitude to the scaled tuneshift which implies
 \beqr
 \rho(\xi_x) & =  & \fr{\rho(\al_x)}{d\xi_x/d\al_x}  =  \fr{1}{2}\exp[-2 \al_x][ \int_0^1   du \; u  [ H_0'(\al_x u) - H_1'(\al_x u) ]^{-1} \\
 H_n(z) & \equiv &   \exp[-z]I_n(z)
 \eeqr 
Hence, using the tuneshift expression in 1D
\beqr
\xi_x(\al_x)  & = &   \int_0^1   du \; [ H_0(\al_x u) - H_1(\al_x u) ] 
\label{eq: xi_alfa} \\
\rho(\xi_x)  & = & \rho(\al_x)[ \fr{\del \xi_x}{\del \al_x}]^{-1} =   2 \exp[ - 2\al_x ]
      [ \int_0^1   du \; u  [ H_0'(\al_x u) - H_1'(\al_x u) ]^{-1} \\
&  \equiv  & 2  \fr{\exp[ - 2 \al_x ]}{ {\rm Jac}(\al_x, \xi_x)}   \label{eq: rho_al}
\eeqr
Here ${\rm Jac}(\al_x, \xi_x)$ is the Jacobian of the transformation from $\al_x \to \xi_x$. 
Eq.(\ref{eq: xi_alfa}) defines $\xi_x$ as a function of $\al_x$.  Inverting this relation (numerically) defines
$\al_x$ as a function of $\xi_x$.  We denote this function $\al_{f}(\xi_x)$.  Inserting this function 
back into Eq.(\ref{eq: rho_al}) yields the functional form
\beq
\rho(\xi_x)  = 2 \fr{ \exp[ - 2 \al_{f}(\xi_x) ] }{{\rm Jac}(\al_{f}(\xi_x))}   \label{eq: rho_xi}
\eeq
In 1D,  the inverse function is straightforward to obtain. Fig.\ref{fig: density1d_tunes} shows plots of the
function
$\xi_x(\al_x)$ and the inverse function $\al_f(\xi_x)$ which resembles a one-sided delta function.

The $p$th moment of the tuneshift i.e.  $\lan \xi_x^0 \ran$ (the norm), $\lan \xi_x^1 \ran, \lan \xi_x^2 \ran,
... $ should
agree in any coordinate system used for the density distribution. We can use this to test the accuracy of
the distribution in $\xi_x$. 
The two lowest moments can be calculated analytically using the known functional forms in terms of $\al_x$.
\beqr
\int_0^1 \rho(\xi_x) d\xi_x & = &  \int_0^{\infty} \rho(\al_x) d\al_x = 2\int_0^{\infty}
 \exp[- 2\al_x] d\al_x = 1 \\
\lan \xi_x  \ran & = & \int_0^{\infty} \xi(\al_x) \rho(\al_x) d\al_x =  2\int_0^{\infty}  \exp[- 2\al_x] d\al_x 
 \int_0^1   du \; [ H_0(\al_x u) - H_1(\al_x u) ] \non \\ 
& = & 2  \int_0^{\infty} d\al_x \; \int_0^1   du \; \exp[- (2 + u)\al_x]  [ I_0(\al_x u) - I_1(\al_x u) ] 
\eeqr
We use the integration result from the table of integrals in  \cite{Grad_Ryzh}
\beq
\int_0^{\infty}e^{- \al z} I_p(\bt z) dz = \fr{\bt^p}{\sqrt{\al^2 - \bt^2}(\al +\sqrt{\al^2 - \bt^2})^p}
\label{eq: integ_gr}
  \eeq
Doing the integral over $\al_x$ first followed by  the integration over $u$, we find
  \beqr
\lan \xi_x  \ran & = & \int_0^{\infty} d\al_x \; \exp[- (2 + u)\al_x]  [ I_0(\al_x u) - I_1(\al_x u) ] \non \\
& = & \int_0^1   du \; \left[\fr{1}{\sqrt{1 + u}} - \fr{1}{\sqrt{1 + u}}\fr{u}{(1 + \sqrt{1 + u})^2}\right] \\
& =  & 4({\rm arcsinh}[1] - \log[2])  = 0.752906
 \eeqr

Numerically the moments can also be calculated using either $\rho(\al_x)$ or using $\rho(\xi_x)$.  The results
 are shown in table \ref{table: moments} where numerical 1 uses the first method and numerical ii
 uses the  second.  The fact that the  moments calculated numerically  with numerical ii agrees exactly with the other methods gives
 confidence that the expression for the density distribution in tuneshift $\rho(\xi_x)$ is correct.

Finally, we calculate the median tuneshift which by definition is the value at which the number of particles is
 the same both above and below the median value $\xi_m$, i.e.
 \beq
 \int_0^{\xi_m} \rho(\xi_x) d\xi_x =  \int_{\xi_m}^{\infty} \rho(\xi_x) d\xi_x = 0.5 \;\;\; \rarw \xi_m = 0.783
 \eeq
The density distribution $\rho(\xi_x)$ along with the average and median tuneshifts are indicated in the
 bottom plot of Fig. \ref{fig: density1d_tunes}. 
\bfig  
\centering
\includegraphics[scale=0.4]{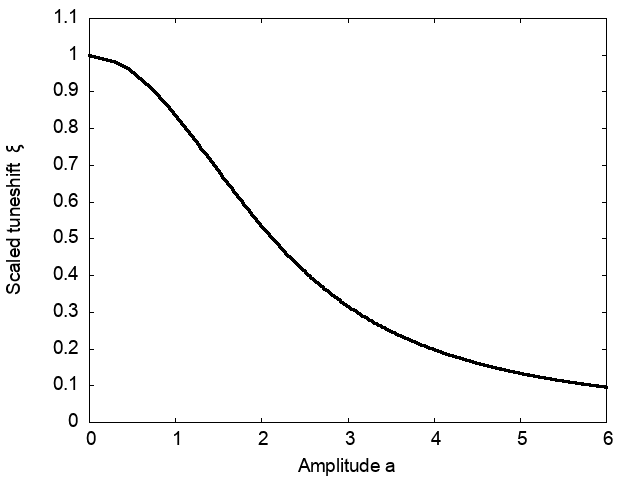}
\includegraphics[scale=0.4]{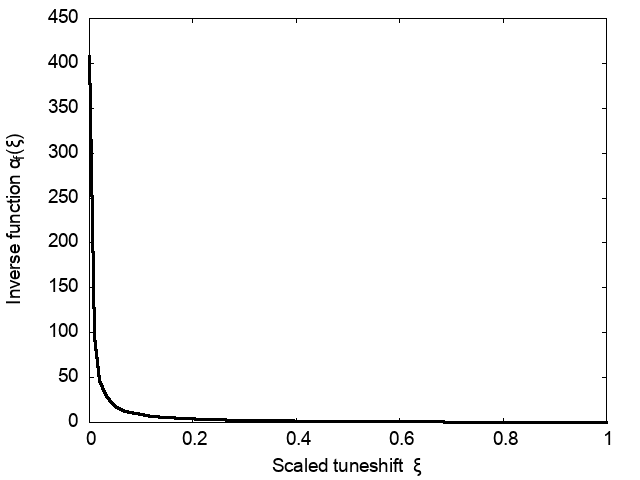}
\includegraphics[scale=0.55]{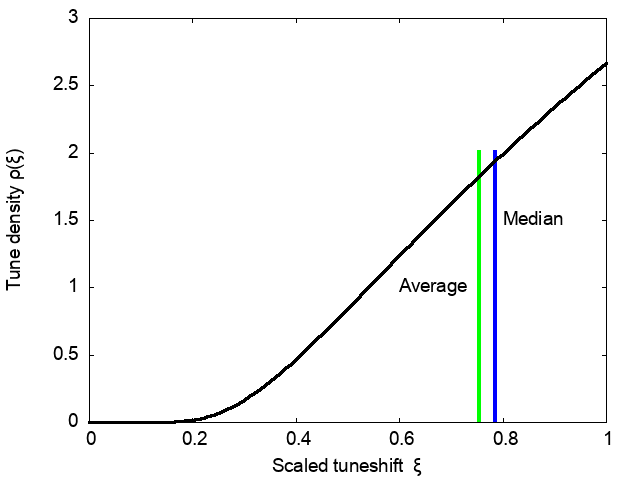}
\caption{Top:  Tune shift $\xi$ as a function of the scaled amplitude $a_x= 2\sqrt{\al_x}$, $a_x$ is the transverse amplitude in units of the rms size; Right: the inverse function $\al_x(\xi)$. Bottom: the density
distribution as a function of the tuneshift.}
\label{fig: density1d_tunes}
\efig  
The 1D density vanishes in the range $a \le \xi_x \le 0.2$ and reaches a maximum at $\xi_x =1$, as expected since that is  the region of maximum density.

\subsection{Density distribution in 2D}

\begin{table}[thb]
  \bec
  \btable{|c|c|c|c|} \hline
  &  analytical   &  numerical i & numerical ii  \\ \hline
   &   \multicolumn{3}{|c|}{ 1D} \\ \hline
normalization &  1.0 & 1.0 &  1.0    \\  
$\lan \xi_x  \ran$  & 0.752906  & 0.752906 &  0.752906 \\ 
$\lan \xi_x^2 \ran$ & n/a &  0.597766    &  0.597766  \\
$\xi_{x, rms}$ & n/a &   0.175782  &  0.175782 \\
 \hline     
    &   \multicolumn{3}{|c|}{ 2D} \\ \hline
normalization &  1.0 & 1.0 &   0.981  \\  
$\lan \xi_x  \ran$  & 0.633389 &  0.633389 &  0.629 \\
$ \lan \xi_x^2 \ran$ & n/a &   0.4293  &  0.4596 \\
$\xi_{x, rms}$ & n/a & 0.1678  & 0.253  \\ \hline
\etable
\caption{Moments of the density distribution calculated in different ways in 1D and 2D.  The two numerical
  ways, labeled as i and ii, describe using the density in amplitude space  $\rho(\al_x, \al_y)$ and
  tune space $\rho(\xi_x, \xi_y)$ respectively.  In all cases numerical i is the more accurate.  }
\label{table: moments}
\eec
\end{table}

The  procedure is the same as in 1D, the details are slightly different because first a nonlinear equation in two
variables has to be solved followed by a 2D interpolation is required.
 Let the density in tune space be  $\rho(\xi_x, \xi_y)$. By the conservation of particle number,
  \beq
  \rho(\al_x, \al_y) d\al_x d\al_y = \rho(\xi_x, \xi_y) d\xi_x d\xi_y
  \eeq
  which implies
  \beq
\rho(\xi_x, \xi_y) =   \rho(\al_x, \al_y) / {\rm Jac}(\xi_x, \xi_y ; \al_x, \al_y)
  \label{eq: rho_al_nu}
  \eeq
  where
  \beq
      {\rm Jac}(\xi_x, \xi_y ; \al_x, \al_y) = \bigg|\bigg| \begin{array}{cc} \fr{\del \xi_x}{\del \al_x}  &  \fr{\del \xi_x}{\del \al_y} \\   \mbox{}  &   \mbox{}  \\
  \fr{\del \xi_y}{\del \al_x}  &  \fr{\del \xi_y}{\del \al_y} \end{array} \bigg| \bigg|    \label{eq: jac}
      \eeq
      The density can be written as
      \beq
      \rho(\al_x, \al_y) =  4 \exp[ - 2\al_x  - 2\al_y]   \label{eq: rhoaxay}
      \eeq
In terms of these variables, the scaled tune shifts are
\beqr
\xi_x(\al_x , \al_y) & = &   
\int_0^1   du \;  [ H_0(\al_x u) - H_1(\al_x u) ] H_0(\al_y u)  \label{eq: xix_a} \\
\xi_y(\al_x , \al_y) & = & 
\int_0^1   du \; H_0(\al_x u)  [ H_0(\al_y u) - H_1(\al_y u) ]   \label{eq: xiy_a}
 \eeqr
The derivatives are found for example as
 \beq
 \fr{\del \xi_x}{\del \al_x}   =  \left(  \int_0^1   du \; u H_0(\al_z u) [ H_0'(\al_x u) - H_1'(\al_x u) ]  H_0(\al_y u) \right)
   \eeq
 These can be used to write the density in tune space using Eqs. (\ref{eq: rho_al_nu}), (\ref{eq: rhoaxay}),
   (\ref{eq: jac}) in terms of $\al_x, \al_y$.
Doing so yields
  \beq
\rho(\xi_x, \xi_y) =  4 \fr{\exp[- 2\al_x - 2\al_y] }{{\rm  Jac}(\xi_x, \xi_y; \al_x, \al_y)}
  \eeq
  The RHS of this equation however is a function of the amplitudes $(\al_x, \al_y)$ while what we want is a
  function of the scaled tune shifts $(\xi_x, \xi_y)$.  That requires an inversion of Eqs. (\ref{eq: xix_a}) and
  (\ref{eq: xiy_a}). It is done in two steps: (1) solving these nonlinear equations to find $\al_x, \al_y$ as
  functions of $\xi_x, \xi_y$ and (2) interpolating these to write these as smooth functions of $\xi_x, \xi_y$.
  Fig. \ref{fig: densities} shows the complete density $ \rho(\xi_x, \xi_y)$ as a function of $\xi_x, \xi_y$
  viewed from two different angles as well the projected density $\rho_x(\xi_x)$ on the $\xi_x$ axis, which is
  obtained   by integrating   over the $\xi_y$ axis, i.e.
  \beq
  \rho_x(\xi_x) = \int_0^1 d\xi_y \;  \rho(\xi_x, \xi_y)
  \eeq

\bfig
\centering
\includegraphics[scale=0.8]{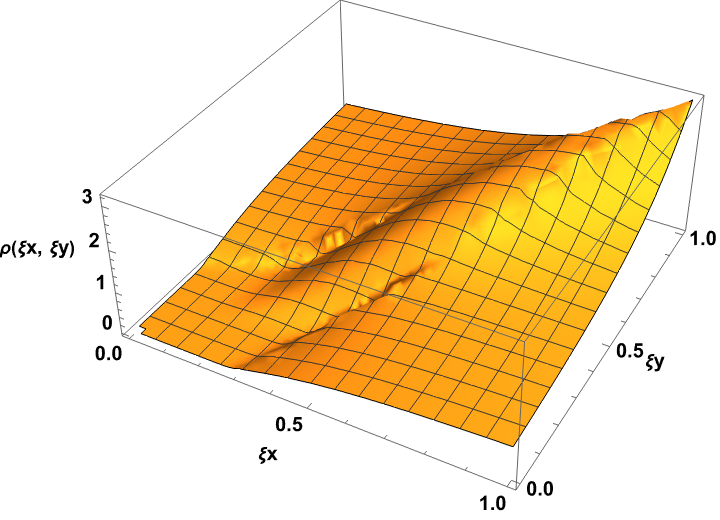}
\includegraphics[scale=0.3]{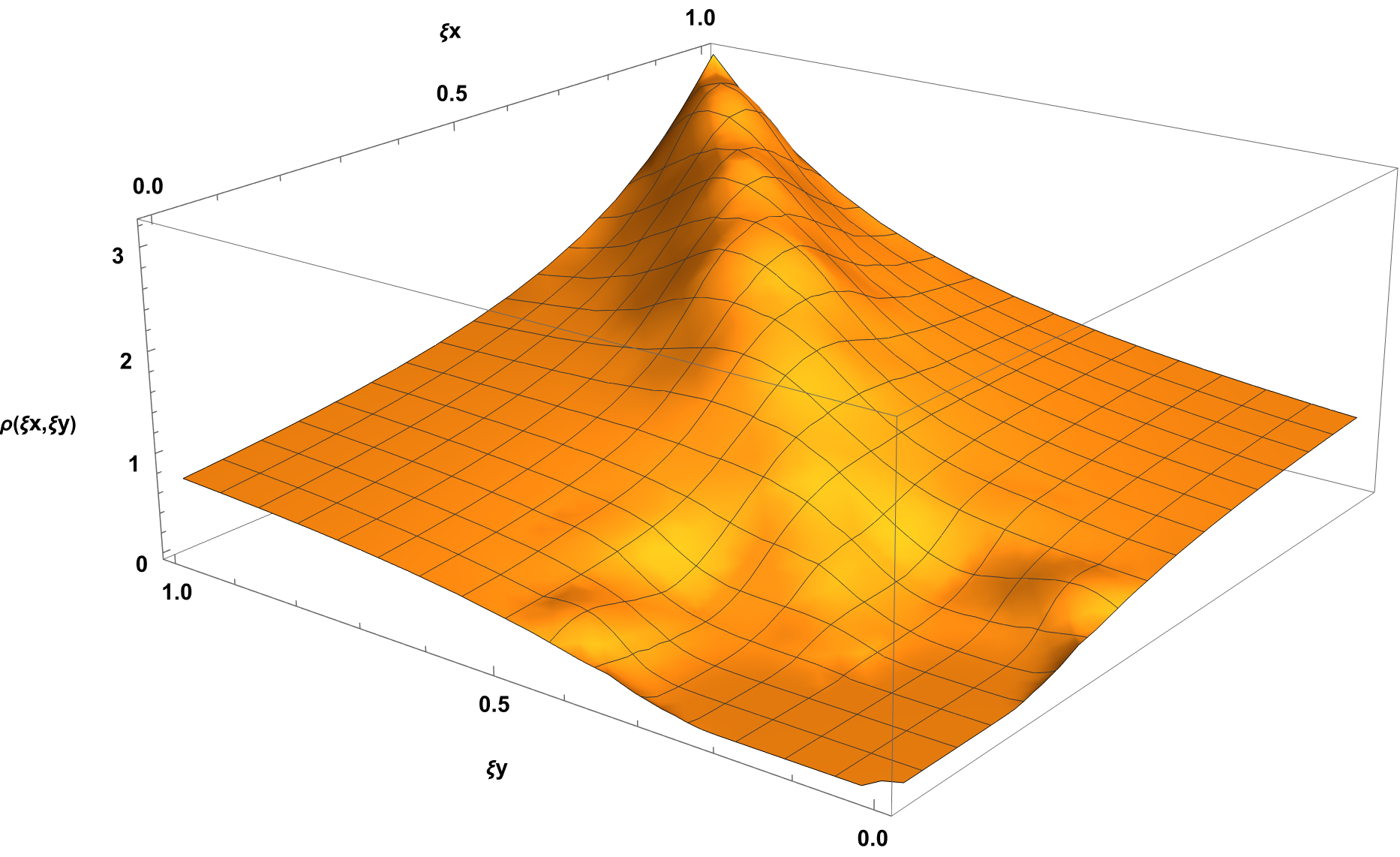}
\includegraphics[scale=0.45]{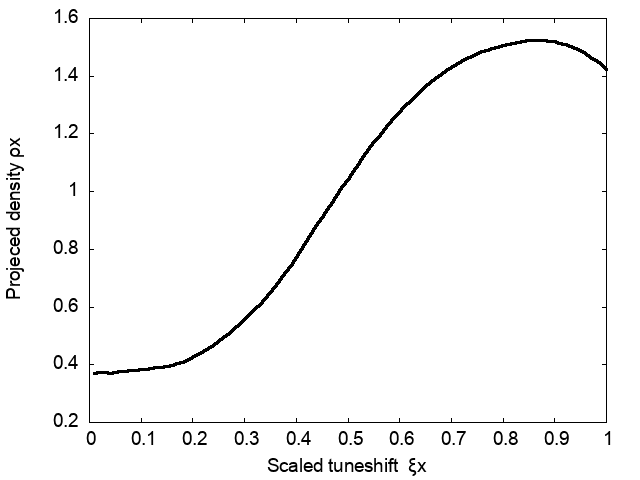}
\caption{Top: density $\rho(\xi_x, \xi_y)$ vs the scaled tune shifts $\xi_x, \xi_y$. 
  Middle:  : this plots shows the density from a different angle, it shows e.g. the density along the $\xi_x$
  and also along the   $\xi_x$ axes. Bottom: the projected   density $\rho_x$ along the $\xi_x$ axis.
  It is non-zero at $\xi_x=0$ and it has a maximum at a value less than $\xi_x=1$.}
\label{fig: densities}
\efig
The density is exactly symmetric along the $\xi_x = \xi_y$  axis, as it must be for round beams. The second plot shows that the density is zero at the origin and we observe that along either the $\xi_x$ axis or the $\xi_y$  axis the density is similar to the 1D density profile seen in
Fig. \ref{fig: density1d_tunes}. The projected density  does not vanish at $\xi_x = 0$ and it has a maximum around $\xi_x = 0.9$ rather than at $\xi_x= 1$.  A little thought shows that is true of any density that has a non-vanishing dependence on both $\al_x, \al_y$.

The moments of the distribution are found as before
  \beqr
 \lan \xi_x  \ran & = & \int_0^{\infty} d\al_x  \int_0^{\infty} d\al_y \; \xi_x(\al_x, \al_y) \rho(\al_x, \al_y) \non \\
& = &  4 \int_0^1   du \; \int_0^{\infty} d\al_x  \int_0^{\infty} d\al_y \; 
  [ H_0(\al_x u) - H_1(\al_x u) ] H_0(\al_y u) \exp[- 2\al_x - 2\al_y] 
 \eeqr
For the integration over $\al_y$, we use the same integral result in Eq.(\ref{eq: integ_gr}) to obtain
 \[ \int_0^{\infty} H_0(\al_y u) \exp[- 2\al_x - 2\al_y]  d\al_y = \fr{1}{\sqrt{1 + u}}  \]
Combining this integration and doing the final integral over $u$, we find
 \beq
 \lan \xi_x  \ran = 4 \ln[ \fr{2 \sqrt{2}}{\sqrt{2} +1} ] = 0.633389
 \eeq
The moments are calculated numerically using both methods used for the 1D distribution, are shown in
  table \ref{table: moments}.  The normalization is not   quite unity with the second method.  
We attribute this $\sim 2\%$ error to numerical issues in the inversion and interpolation required to find the
functions $\al_x(\xi_x, \xi_y),  \al_y(\xi_x, \xi_y)$. This can be used to find the correct moments by dividing the
raw moments by the normalization. With this correction, the error in the first moment is $\sim 0.7\%$ while
the error in the second moment is 7\%.

We briefly consider application of these results to beam stability with octupoles and space charge, a more detailed
study will be reported elsewhere. Typically the stability  is considered using the dispersion relation with a 2D
betatron spread from both octupoles and space charge. This relation was used to derive stability curves
\cite{Met_Rugg} with a parabolic transverse density which results in  a space charge tune spread linear in  the
actions that matches the octupolar tune spread dependence. It may be possible to use the exact space charge
tune spread
$\Dl\nu_{x, sc}(a_x, a_y) = \Dl\nu_{x, sc} \xi_z(a_x, a_y)$ to obtain stability curves for Gaussian bunches. The
extension to 3d stability is in principle straightforward, using the 3d space charge tuneshifts
$\xi_{x, y}(a_x, a_y, a_z)$ found in section \ref{Sec: tunes_theory}. A more direct  use of the density curves
derived in this section would be to
check the results derived from PIC simulations against the exact results obtained here. 
We also note that the rms tune spread calculated in table \ref{table: moments} may be directly related to the
decoherence time following a kick; this time in 2D is longer than the typical $1/\Dl\nu_{sc}$ time scale associated with 1D decoherence.

\clearpage

\section{Application to IOTA} \label{Sec: IOTA} 

\begin{table}
  \bec
  \btable{|c|c|} \hline
  \multicolumn{2}{|c||}{IOTA proton parameters} \\ \hline
Circumference & 39.97 [m] \\
 Kinetic energy & 2.5 [MeV] \\
Maximum bunch intensity /current & 9$\times 10^{10}$ / 8 [ma] \\
Transverse normalized rms emittance & (0.3, 0.3) [mm-mrad] \\
Betatron tunes & (5.3, 5.3) \\
Natural chromaticities &  (-8.2, -8.1)     \\ 
Average transverse beam sizes (rms) &  (2.22, 2.22) [mm] \\ 
Kinematic $\gm$ / transition $\gm_t$ &  1.003 / 3.75 \\
Rf voltage & 400 [Volts]  \\
Rf frequency  /  harmonic number & 2.2 [MHz]   / 4   \\
Bucket length &  $\sim 10$ [m] \\
Bucket half height in $\dl p/p$ &   3.72 $\times 10^{-3}$ \\
Rms bunch length &  1.7 [m] \\
Rms energy /momentum spread &  1.05$\times 10^{-5}$ / 1.99 $\times 10^{-3}$   \\
\hline
\etable
\caption{Machine and beam parameters of the IOTA proton ring}
\label{table: parameters}
\eec
\end{table}

IOTA is an accelerator that was designed to test the concept of nonlinear integrable lattices \cite{Dan_Nag}. 
The R \& D program with electrons and protons was discussed in \cite{Antipov}. The ring has been operated
with electrons since commissioning began and several notable results  have been achieved, including the
demonstration of optical stochastic cooling \cite{Jarvis}. Proton operation is scheduled to begin in 2024 when the concept of
achieving high space charge tune shifts with a nonlinear integrable lattice will be tested.

In this section, we will evaluate the space charge footprints theoretically and compare with particle tracking. 
This is done in an otherwise completely linear lattice; we note that emittance growth and beam loss was
studied in a partially integrable lattice with octupoles in \cite{David_21}. 
Table \ref{table: parameters} shows the relevant parameters of the iota proton ring.
First, we consider whether the beam is sufficiently round everywhere in the ring for the round beam
expressions for the footprint to be applicable. Fig. \ref{fig: sigyx} shows the ratio of the vertical to horizontal
beam sizes along the ring. The ratio varies between 0.5 to 5.0 with a mean value of 1.2. The mean value may
not be relevant here, as the fluctuations are fairly large. We therefore use the general expression for the
tune shifts but we also compare with the round beam forms as well as an approximation discussed in section
\ref{Sec: tunes_theory}. 
\bfig
\centering
\includegraphics[scale=0.5]{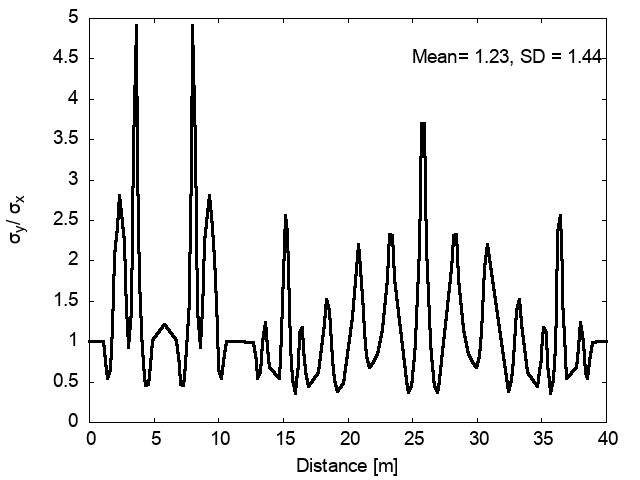}
\caption{ratio of $\sg_y/\sg_x$ around the ring.}
\label{fig: sigyx}
\efig
We discuss first the theoretical footprints under the different assumptions discussed in section \ref{Sec: tunes_theory}.
In the limit of long bunches, which is  valid for IOTA, we can use Eq.(\ref{eq: dnux_gen}).
This involves calculating the tune shift at each longitudinal location and averaging over the locations. This method
requires doing an integration at each location. We can reverse the order and instead do the averaging first and do
a single integration instead.
\beqr
\Dl\nu_x(a_x, a_y, a_z) & = &   \Dl\nu_{x, sc} \int_0^1   du \;  \left\{ {\big \lan}  
  \exp[- \fr{ a_z^2 u}{4}]  I_0\left(\fr{ a_z^2 u }{4 }\right)
 \exp[- \fr{ a_x^2 u}{4}]\left[ I_0(\fr{ a_x^2 u}{4}) - I_1(\fr{ a_x^2 u}{4})\right] 
 \right.  \non   \\
& & \left. \times   \exp[- \fr{ a_y^2 u}{4}]  \left[\fr{1}{[( \sg_y^2/\sg_x^2 - 1 )u + 1 ]} \right]^{1/2}   
 I_0\left(\fr{ a_y^2}{4}\fr{u}{(1 - \sg_x^2/\sg_y^2 )u + \sg_x^2/\sg_y^2}\right)
{\big  \ran_s}   \right\}  \non \\
\eeqr
This reduces the time required for evaluation and we found that the differences in numerical values are negligible,
at least in the case of IOTA. 

The left plot in Fig. \ref{fig: ftprints_theory} shows a comparison of the footprints based on the general expression in
Eq.(\ref{eq: dnux_gen}), and that based on the round beam expression in Eq.(\ref{eq: dnux_long_round}). 
The differences are small; this is to be expected as iota has been designed to have axial symmetry almost
everywhere in the ring in order to preserve integrability \cite{Dan_Nag}. The right plot in this figure shows
the footprints with synchrotron oscillations at two amplitudes $a_s = 1, 2 $.
As expected from Fig. \ref{fig: facz}, the total tuneshift at $a_s =1$ is about 90\% and at $a_s=2$, the total
tuneshift is 70\% of the value at $a_s=0$.
\bfig
\centering
  \includegraphics[scale = 0.425]{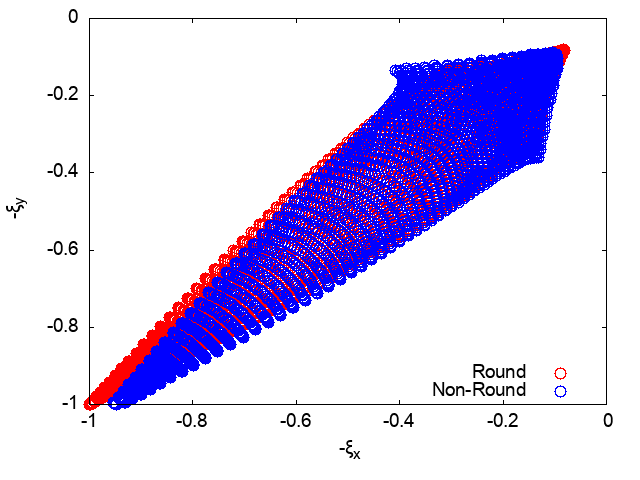}
\includegraphics[scale = 0.425]{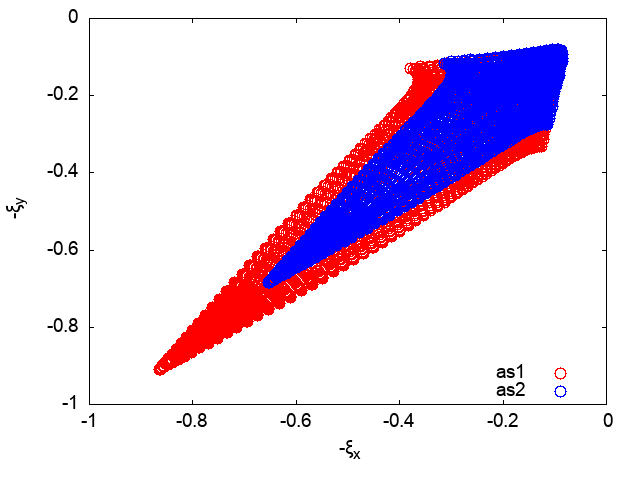}
  \caption{Left: comparison of round and non-round footprints without synchrotron oscillations. Right:
    footprints with synchrotron oscillation amplitudes of 1$\sg_s$ (in red) and 2$\sg_s$ (blue). }
\label{fig: ftprints_theory}
  \efig

  The theory of the amplitude dependent tuneshifts assumes that the particles stay at constant amplitudes while
  executing betatron oscillations. This is not always true, especially at high intensities. We examine this assumption
  by testing
emittance growth with pyorbit simulations \cite{pyorbit}. Many details on the pyorbit simulations and
their validation
can be found in earlier reports \cite{Runze_20, David_21}. Detailed analysis had shown that  there was
good agreement between theory and simulations in all the tested regimes. In the simulations reported here,
all machine nonlinearities were turned off, space charge was the only nonlinearity.
The results reported in \cite{Runze_20} showed that initial beam losses can be minimized by a process of
slow initialization in which the charge per macroparticle is increased over about 40 turns to full value and the
beam was injected into a lattice that was rms matched to equilibrium beam sizes. The PIC parameters that
led to convergence were found to have the following values: number of macroparticles = $5 \times 10^5$,
grid size = 128 x 128 x 5, and the number of space charge kicks per betatron wavelength =  63. These
values were used in the simulations discussed below.

It is not completely straightforward to compare space charge simulations with theory especially at high
intensities. 
The simplest is to compare the space charge tune shifts. The complications arise from two effects
theory assumes that the tune shift is calculated at constant emittances, which is not the case as the space
charge increases. The second complication is related to PIC simulations. It has been observed that this method
causes orbits to be chaotic at small amplitudes close to the origin \cite{Schmidt}.  We dealt with the first
issue by using the average emittance over the time used for the tuneshift calculation.
For the second issue, we distribute particles over 100 different angles at the same
small amplitude and average over the angles to reduce the fluctuations in the tune shift  value. 
Another complication at high intensities is that because 
 The FFT aliases tunes to be in the range 0 - 0.5, it cannot determine if the tunes are below or above the
 half integer   when the space charge tune shift exceeds 0.5. Here we determined the correct tuneshifts by selecting
 the value that increased with intensity, without deeming the integer parts.
Previously we had determined the complete tune (integer and fractional parts) with the alternative
  method of counting the  number of betatron oscillations over a thousand turns and found that the simulated
  tunes calculated both ways agreed well with each other and with theory \cite{Runze_20}.
  
At a low intensity of $10^9$ particles/bunch, there are only
  fluctuations due to numerical noise in the PIC simulations. These are observed to be around 0.4\%, which is
  close to the expected level $\sim 1/\sqrt{n} = 0.14$\% with $n = 5 \times 10^5$. 
\begin{figure}%%
  \centering
  \includegraphics[scale = 0.55]{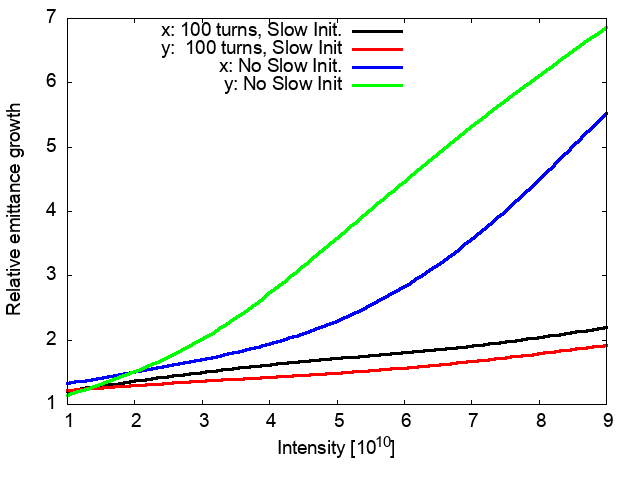}
  \caption{Relative emittance growth over 1000 turns as a function of the intensity for two conditions:
    without slow initialization and   with a slow initialization of 100 turns.}
 \label{fig: emitgrowth}
 \efig
Fig. \ref{fig: emitgrowth} shows the emittance change over a range of intensities;  the last value
  $ 9\times 10^{10}$   corresponds to the maximum design bunch intensity.
Since there is very little  observable emittance growth at the lowest intensities in this range,
we expect the simulated tunes to be close to theoretical values. 
However, the tune shifts are too small to be accurately computable with an FFT over $\sim$1000 turns,
  especially to resolve the tune-shift for neighboring particles. At intermediate intensity of 10$^{10}$, there is a  larger   emittance growth pf $\sim 10\%$  while at $9\times 10^{10}$, the emittance grows by nearly a factor of  ten without slow initialization.
\bfig
\centering
  \includegraphics[scale = 0.425]{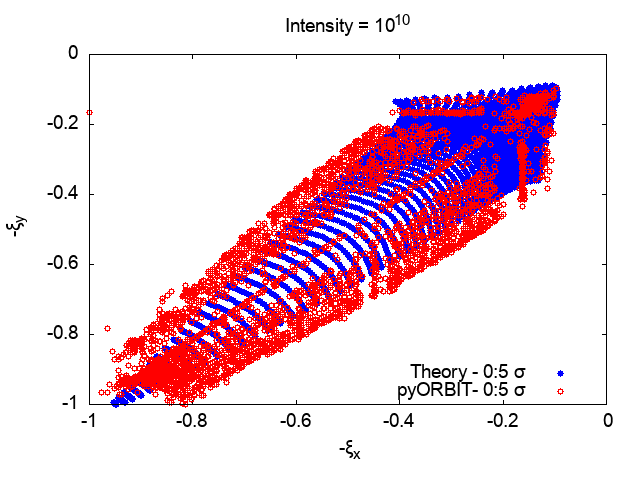}
  \includegraphics[scale = 0.425]{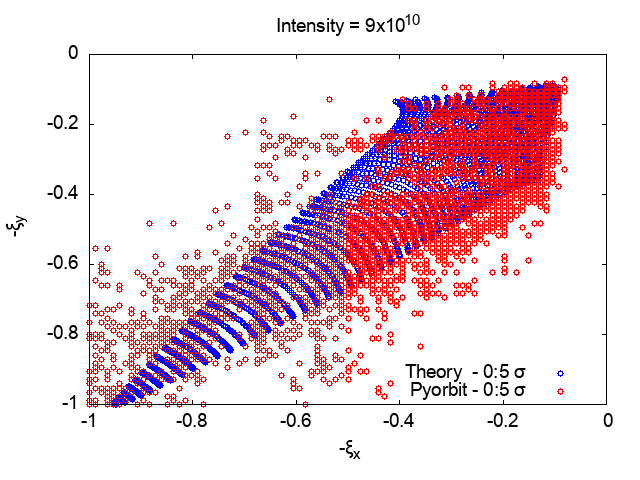}
  \caption{Footprints from pyorbit tracking and theory with non-round beams. Left: intensity= 10$^{10}$,
Right: intensity= $9 \times 10^{10}$}
\label{fig: footprints-sim_theor}
\efig 
The tune footprints are calculated using $5\times 10^5$ macroparticles and using 5000 test particles
distributed transversely from 0 - 5$\sg$ and with zero synchrotron amplitude. Fig. \ref{fig: footprints-sim_theor}
shows the footprints at 10$^{10}$ and $9\times 10^{10}$ intensities obtained with pyorbit simulations and
compared with the theoretical values using Eq.(\ref{eq: dq_Long_NR}). In both cases, we have scaled the
numerical footprints by the maximum tuneshift, so the analytical and numerical footprints can be easily compared.
At the lower intensity, the two footprints agree reasonably well although the simulated footprint is wider at
amplitudes from 1 to 5$\sg$. At the intensity of $9\times 10^{10}$, the simulated footprint is even wider and
the agreement is not as close, which is to be expected. The single most important reason for the increasing
discrepancy is that the theory assumes that all particles move on invariant actions which is not true at high
intensities. Nevertheless, the theoretical footprint can be useful both as a benchmark tool and also to
quickly determine the important resonances that can be crossed by the footprint at a chosen working point.

%\clearpage

\section{Conclusions} 

We derived tune shifts with amplitude in terms of a universal dimensionless parameter under quite general
  conditions that are valid for space charge or beam-beam interactions. We included multiple interaction points and synchrotron oscillations. Our focus is on space charge interactions mainly and the inclusion of multiple
  interactions as well as beams with arbitrary transverse aspect ratios is especially important. 
We then used the analytical tune shifts to derive semi-analytical expressions for the density distribution of tunes
  assuming that the density is a Gaussian function of the phase-space coordinates. The tune distribution requires
  an inversion of the functional arguments followed by a numerical interpolation. We emphasize that the
  tune distribution thus obtained requires no numerical simulations. This is important because the density at the
  maximum tune shift requires very high sampling of this region and quite often the simulations get the wrong
  shape of the density in this region. The density is expressed in terms of variables $(\xi_x, \xi_y)$ which are the
  tune shifts scaled by the maximum  tune shifts. Therefore the density $\rho(\xi_x, \xi_y)$ has the same form and
  shape for both space charge and beam-beam interactions. 
With the method presented here, we verified that the low order moments
  of the distribution are preserved in the transformation; exactly in 1D and with a $\sim 2\%$ error in 2D for the
  zeroth moment, see table \ref{table: moments} for the other moments.  These error could be further reduced by
  improving the numerical schemes for the function inversion and interpolation. This calculation of the density
  distribution in tunes will enable a more accurate modeling of landau damping with space charge and an
  external nonlinearity such as octupoles as well as the damping with beam-beam interactions. 

We checked the tune spread calculations for the IOTA proton ring  with simulations using the pyorbit code.
At the highest intensities planned with bunched beams, there is substantial emittance blow up and steps will
  need to be taken to mitigate emittance growth and beam loss. We used a numerical scheme of slow
  initialization to reduce the growth and prevent beam loss over the short time scale of the simulation.
Using this scheme, we found generally good agreement between the footprints calculated by theory and
  simulation. The expressions for the theoretical footprints developed in this paper should therefore be useful
  for bench-marking other space charge simulation codes as well as determining working points relatively free of
  low order space charge driven resonances.

\vspace{2em}
\noi {\bf \large Acknowledgments} \\
I thank former undergraduate interns David Feigelson (U Chicago) and Runze Li (UW, Madison; now at Yale) and
colleague Francois Ostiguy for their enthusiastic collaboration on a project to model space charge effects in IOTA.
I am especially thankful to Runze for writing a  library of pyorbit codes needed for modeling IOTA; these codes are
available at his github site \cite{Runze_github}. 

The work has been supported by the Fermi Research Alliance, LLC under Contract No. DE-AC02-07CH11359 with
the U.S. Department of Energy, Office of Science, Office of High Energy Physics.

\end{document}